\documentstyle [preprint,aps,epsf,psfig]{revtex}
\begin{document}
\title{The Stripe-Phase Quantum-Critical-Point Scenario \\ 
for High-$T_c$ Superconductors}
\date{preprint: \today}
\author{
S. Caprara, C. Castellani, C. Di Castro, M. Grilli, A. Perali and M. Sulpizi
}
\address {Dipartimento di Fisica - Universit\`a di Roma ``La Sapienza'' \\
and Istituto Nazionale di Fisica della Materia, Unit\'a di Roma 1 \\
P.le A. Moro 2, I-00185 Roma - ITALY}
\maketitle
\begin{abstract}
A summary is given of the main outcomes of the 
quantum-critical-point scenario for high-$T_c$ superconductors, developed 
in the last few years by the Rome group. Phase separation, which commonly 
occurs in strongly correlated electronic systems, turns into a stripe 
instability when Coulomb interaction is taken into account. 
The stripe phase continuously 
connects the high doping regime, dominated by charge degrees of freedom to the 
low doping regime where spin degrees of freedom are most
relevant. Dynamical stripe 
fluctuations enslave antiferromagnetic fluctuations at high doping.
Critical fluctuations near the stripe instability
mediate a singular interaction between quasiparticles,
which is responsible for the non-Fermi liquid behavior in the metallic phase 
and for the Cooper pairing with $d$-wave symmetry in the 
superconducting phase. 
\end{abstract}

\vspace{0.2cm}

\section{The framework}

Since the discovery of superconducting (SC) copper oxides \cite{BM}
a formidable effort has been produced to provide a unified theory for 
the rich phase diagram of these materials (Fig. 1).

The antiferromagnetic (AFM)
phase at zero and very low doping is usually described
as resulting from the strongly correlated nature of the copper-oxygen
planes, within the Hubbard model or the related $t-J$ model.

As far as the SC phase is concerned the main points
under investigation are the nature of the (strong) pairing mechanism,
the unusual ($d$-wave) symmetry of the order parameter and the
strong dependence of the critical temperature $T_c$ on the doping $x$.

The properties of the normal 
state are to some extent even more challenging, the standard
Fermi liquid (FL) theory appearing to be violated.
The copper oxides are characterized by
a low dimensionality, revealed by the strong anisotropy
of the transport properties. In the metallic phase above $T_c$ at
optimum doping a non-FL behavior sets in,
with a linear in-plane resistivity over a wide
range of temperatures \cite{T}, 
indicating the absence of any energy scale, besides
the temperature itself. In the underdoped region two new temperature scales
appear above $T_c$. The higher, $T_0(x)$,
marks the onset of a new regime characterized by a reduction of the 
quasiparticle
density of states, and is mainly revealed by the presence of broad maxima in 
the spin 
susceptibility \cite{CHI}, and by a 
downward deviation of the in-plane resistivity
as a function of the temperature \cite{RHO}.
At a lower temperature $T^*(x)$  a (local) gap in the spin and charge
channels appears in ARPES \cite{marshall,harris,ding}, 
NMR \cite{berthier}, neutron scattering 
\cite{neutronrm,neutronmo,neutronma,neutronpe} and
specific heat measurements \cite{loram}.

Anderson proposed \cite{AND} to extend the $d=1$ Luttinger Liquid
behavior to $d=2$ and explain the anomalies in the metallic phase.

However, no sign of such a new quantum metallic state 
was found within a 
renormalization-group approach in $d=2$ \cite{rg2d}. Rather a dimensional
crossover drives the system to a FL state as soon as $d>1$
in the presence of short-range forces \cite{CDM}. When long-range forces
are taken into account a non-FL behavior may arise in the
presence of a sufficiently singular interaction $\Gamma_{eff}(q)\sim
1/q^\alpha$ with $\alpha\ge 2(d-1)$ \cite{MAC}. 

The onset of an instability is a
mechanism which provides a suitable
singular scattering. Indeed
critical fluctuations mediate an effective interaction between
quasiparticles
$\Gamma_{eff}(q,\omega)\simeq -V/[(q-Q)^2+\kappa^2-{\rm i}\gamma
\omega]$ where $V$ is the strength of the static effective potential at
criticality, $Q$ is the critical wavevector, $\kappa^2\sim
\xi^{-2}$ is a mass
term which is related to the inverse of the correlation length 
and provides a measure of the distance
from criticality. The characteristic time scale of the
critical fluctuations is $\gamma$. 
We point out that the static part of this
effective interaction has the form of the Ornstein-Zernike critical
correlator.

Proposals about the nature of the relevant instability 
include ($ i$)
an AFM Quantum Critical Point (QCP) \cite{sy,pines}, ($ ii$) a charge-transfer 
instability \cite{varma}, ($ iii$) an ``as-yet unidentified''
QCP regulating a first-order phase-transition
between the AFM state and the SC state \cite{laughlin}, ($ iv$)
an incommensurate charge-density-wave (ICDW) QCP \cite{cdg1,cdg2}.

The theory of the AFM QCP \cite{pines}
is based on the hypothesis that the presence
of an AFM phase at low doping is the relevant feature common to
all cuprates and on the observation that strong AFM
fluctuations survive  at larger doping
\cite{neutronrm,neutronmo,neutronma,neutronpe}. However at doping as high as
the optimum doping it is likely that charge degrees of freedom play a major role
whereas spin degrees of freedom {\sl follow} the charge dynamics, 
and are enslaved \cite{tranq}
by the charge instability controlled by the ICDW QCP
\cite{cdg1,cdg2}. The AFM fluctuations
are thus extended to a region far away from the AFM QCP, due to the
natural tendency of hole-poor domains towards antiferromagnetism.
The strong interplay between charge and spin degrees of freedom gives rise to
the ``stripe phase'' which continuously
connects the onset of the charge instability (ICDW QCP) at
high-doping with the low-doping regime
characterized by the tendency of the AFM background to expel mobile holes.
Because of this we shall more properly refer to the ICDW QCP as the stripe
QCP. Therefore
we point out that the presence of a 
stripe QCP is not alternative to the presence
of the AFM QCP, which is found at lower doping. 
The two points
control the behavior of the system in different regions of doping. On 
the other hand the existence of a QCP at optimum doping,
where no other energy scale besides the temperature is
present in transport measurements, is the natural
explanation for the peculiar nature of this doping regime in the phase
diagram of all SC copper oxides.

There is an increasing
amount of theoretical and experimental evidence in favour of the
presence of a QCP near optimum doping.
Indeed, the instability with respect to
phase separation (PS)
into hole-rich and hole-poor regions is a generic
feature of models for strongly correlated electrons with
short-range interactions \cite{ps}, which is turned into a frustrated
PS \cite{ek} or in an ICDW instability \cite{cdg1}
when long-range Coulomb forces are taken into account to guarantee large-scale
neutrality. Close to PS (or ICDW)
there is always a region in parameter space where Cooper pair formation is
present, pointing towards a connection between PS
and superconductivity. 

The most compelling evidence for a QCP near optimum doping
is provided by the resistivity measurements. An insulator-to-metal transition
is found when the SC phase is suppressed by means of
a pulsed magnetic field \cite{boebinger}.
When extrapolated to zero temperature
such a transition takes place near optimum doping, and anyway at
too high a doping to be associated to the spin-glass 
region \cite{chou}
characterized by the local moment formation as seen in the
muon experiments \cite{musr}. The spin-glass region 
should be instead a signature of the
coexistence of superconductivity with antiferromagnetism proper of the 
$SO(5)$ theory \cite{zhang}.
Moreover a clear indication that this insulator-to-metal transition 
\cite{boebinger} is 
driven by some spatial charge ordering is provided by its occurrence 
at a much higher temperature in samples near the filling
1/8. Commensurability effects near this ``magic'' filling
have repeatedly been reported in related compounds \cite{tranquada}.

Hints for a critical behavior of the charge susceptibility 
come from the study of the chemical potential shift in PES and BIS 
experiments \cite{ino}. A dramatic flattening of the $\mu$ vs $x$ curve
starting at $x\simeq 0.15$ could be the signature of a 
divergent compressibility.
Finally, stripes of either statical or dynamical nature are seen 
in neutron scattering experiments \cite{tranq}, EXAFS \cite{exafs} and X-ray 
diffraction \cite{ics}.

It must be pointed out that the characteristics of the 
stripe phase produced by the ICDW instability are
system and model dependent. The direction of the
critical wavevector $Q_c$ is diagonal in YBCO \cite{neutronmo},
and in nickelates \cite{ni} 
where one-hole filled domain walls are present, 
and vertical in Nd doped LSCO
\cite{tranquada}, with
half-filled domain walls. It has been shown that a strong on-site Hubbard
repulsion and a long-range potential stabilize vertical half-filled
stripes \cite{goetz}.

The stripe-QCP scenario provides therefore a scheme to interpolate between 
the repulsion which gives
rise to the AFM state at low doping and the attraction giving rise
to SC, through the (local) PS or ICDW. 

\section{The normal state}

To investigate the effect of the stripe QCP on the normal-state properties
of the system we are led to consider
an effective interaction between quasiparticles 
\begin{equation}
\Gamma_{eff}(q,\omega)=-\sum_{i=c,s}
\displaystyle{ V_i\over [\kappa_i^2+\eta_{q-Q_i}
-{\rm i}\gamma_i \omega]
},
\label{eff}
\end{equation}
where $\eta_q=2-(\cos q_x+\cos q_y)$, which is
mediated by both
charge ($c$) and (enslaved) spin ($s$) fluctuations. The $q$ dependence
in (\ref{eff}), was taken in the cos-like form
to reproduce the $(q-Q_i)^2$ behavior close to the critical
wavevectors $Q_i$ and to maintain the lattice periodicity near the zone
boundary. 

We point out that the above form (\ref{eff}) for the
interaction mediated by charge fluctuations was found within a slave-boson 
approach to the Hubbard-Holstein model with long-range Coulomb
interaction, close to the ICDW instability \cite{becca}. 
The same form 
mediated by spin fluctuations corresponds to
the dynamic susceptibility proposed by 
Millis, Monien and Pines \cite{millis} to fit NMR and neutron
scattering experiments, in the limit of strong damping. 

We take a free-electron spectrum of the form
\begin{equation}
\xi_k=-2t(\cos k_x +\cos k_y)-4t'\cos k_x \cos k_y -\mu,
\label{xik}
\end{equation}
where nearest-neighbor 
($t$) and next-to-nearest-neighbor ($t'$) hopping terms
are considered, to reproduce the main features of the band dispersion
and the Fermi surface (FS) observed in SC copper oxides. 
The chemical potential is
treated self-consistently within a perturbative approach, to fix
the number of particles.

The first-order in perturbation theory yields an
electron self-energy
$\Sigma(k,\varepsilon)=\sum_{i=c,s} \Sigma_i(k,\varepsilon)$, where
$$
{\rm Im}\Sigma_i(k,\varepsilon)=
{V_i\over \gamma_i}\int \int_{-\pi}^{+\pi} 
{dk_x^\prime dk_y^\prime\over 4\pi^2}
\left[f(\xi_{k'})+b(\xi_{k'}-\varepsilon)\right]\times
$$
\begin{equation}
\displaystyle{
{\varepsilon -\xi_{k'} \over [\varepsilon -\xi_{k'}]^2 +\Omega_i^2(k-k')}
},
\label{imsigma}
\end{equation}
$\Omega_i(q)=\gamma_i^{-1}[\kappa_i^2+ \eta_{q-Q_i}]$,
$f(\varepsilon)$ is the Fermi function and $b(\varepsilon)$ is the Bose
function. The real part of the self-energy ${\rm Re}\Sigma(k,\varepsilon)$
is obtained by a Kramers-Kr\" onig transformation of (\ref{imsigma}). 
To keep the inversion symmetry $k\to -k$
we symmetrize the self-energies $\Sigma_{c,s}$ with respect to $\pm Q_{c,s}$.
We assume that $Q_s=(\pi,\pi)$, neglecting the possibility for a
discommensuration of the spin fluctuations in a (dynamical) stripe phase 
\cite{tranq}. This would introduce minor changes in our results. The
relevant direction of the critical wavevector $Q_c$ is 
still debated and can be material dependent \cite{mook}.
We analyze here the case $Q_c\sim (1,-1)$ 
which has been suggested by the analysis of ARPES experiments on
Bi2212 \cite{bianconi}.

In the absence of superconductivity the stripe QCP, located at $\kappa_c^2=0$,
is the end point of two lines which divide the $T$ vs $x$ plane
in three regions, the ordered-phase region at lower doping and 
low temperature, the quantum disordered region at higher doping and low
temperature and the quantum critical region around the critical
doping $x_c\simeq x_{optimal}$ where the only energy scale is
$\kappa_c^2\sim T$ and the maximum violation of the FL behavior
in the metallic phase is found. In this region the system is characterized
by an anomalously large quasiparticle inverse scattering time
$1/\tau_k\sim \sqrt{T}$ at those point of the FS connected
by the critical wavevectors (hot spots) and displays a linear-in-$T$
resistivity in $d=2$
\cite{hr}, which is turned to a $T^{3/2}$ which could be a 
signature of a $d=3$ quantum critical behavior for less anisotropic systems,
with the change in the temperature dependence occurring over the whole
temperature range by increasing doping.

To study the effect of the singular interaction (\ref{eff}) on the
single-particle properties in the metallic phase we calculate
the spectral density $A(k,\varepsilon)=\pi^{-1}|{\rm Im}\Sigma(k,\varepsilon)| 
/ \{ [\varepsilon-\xi_k-{\rm Re}\Sigma(k,\varepsilon)]^2+[{\rm Im}\Sigma
(k,\varepsilon)]^2 \}$. To allow for a comparison
with ARPES experiments we analyze the
convoluted spectral density
\begin{equation}
\tilde{A}(k,\varepsilon)=\int_{-\infty}^{+\infty}d\varepsilon'
A(k,\varepsilon')f(\varepsilon') E(\varepsilon'-\varepsilon)
\label{convak}
\end{equation}
which takes care of the absence of occupied states
above the Fermi energy,
through the Fermi function $f(\varepsilon)$, and of the experimental
energy resolution $\Delta$, through a resolution function 
$E(\varepsilon;\Delta)$. We take $E(\varepsilon;\Delta)=
\exp(-\varepsilon^2/2\Delta^2)/\sqrt{2\pi\Delta^2}$ or
$=[\vartheta(\varepsilon+\Delta)-\vartheta(\varepsilon-\Delta)]/2\Delta$
according to numerical convenience. For the sake of
definiteness we choose our parameters in (\ref{xik})
to fit the band-structure
and FS of Bi2212, namely
$t=200$ meV, $t'=-50$ meV, and $\mu=-180$ meV, 
corresponding to a hole doping $x\simeq 0.17$ with respect to half filling.
The parameters appearing in the effective interaction (\ref{eff})
where taken as $V_{c,s}=400$ meV, $\kappa_{c,s}= 0.01$ and
$\gamma_{c,s}=0.01$ meV$^{-1}$.

The quasiparticle spectra are characterized by a coherent quasiparticle
peak at an energy $\varepsilon\simeq \xi_k$ and by shadow peaks
at energies $\varepsilon\simeq \xi_{k-Q_i}$, produced by the interaction
with charge and spin fluctuations. The shadow peaks do not generally
correspond to new
poles in the electron Green function and are essentially incoherent, although
they {\sl follow} the dispersion of the shadow bands. Their 
intensity varies strongly with $k$ and increases when $\xi_{k-Q_i}$
approaches the value $\xi_k$.
In particular at the hot spots, where $\xi_k\simeq\xi_{k-Q_i}\simeq 0$ and
the non-FL inverse scattering time $1/\tau_k \sim \sqrt{\xi_k}$ \cite{hr}, 
there is a suppression of the coherent spectral weight at the Fermi energy.

We also study the $k$-distribution of low-laying spectral weight
$w_k={\tilde A}(k,\varepsilon=0)$.
The transfer of the
spectral weight from the main FS to the different branches of the
shadow FS at $\xi_{k-Q_i}\simeq 0$ produces features which
are characteristic
of the interaction with charge and spin fluctuations and of their
interplay. In particular
the symmetric suppression of spectral weight at the $M$ points
of the Brillouin zone, which would be due to spin fluctuations alone,
is modulated by charge fluctuations (Fig. 2). This is also the case for the
(weak) hole pockets produced by spin fluctuations
around the points $(\pm \pi/2,\pm \pi/2)$. The interference with the branches 
of the shadow FS due to charge fluctuations enhances these pockets 
around $\pm(\pi/2,\pi/2)$
and suppresses them around $\pm(\pi/2,-\pi/2)$ (Fig. 2).
Experimental results on this issue are controversial. Strong shadow peaks
in the diagonal directions, giving rise to hole pockets in the FS,
have been reported in the literature \cite{bianconi,larosa}, where
other experiments found only weak (or even absent) features \cite{marshall}.

We point out that, because of the transfer of spectral
weight to the shadow FS, the experimentally observed FS may be rather
different from the theoretical FS, determined as $\xi_k+{\rm Re}\Sigma
(k,\varepsilon=0)=0$. The observed evolution of the FS could, indeed,
be associated with the change in the distribution of the low-laying 
spectral weight, without the topological change in the quasiparticle FS which
was proposed in \cite{chubukov}.

\section{superconductivity}

In the stripe-QCP scenario
the dynamical precursors of the ICDW mediate an attractive interaction
in the Cooper channel \cite{perali}. 
As a matter of simplification we solve
the BCS-like equation
$$
\Delta_k=-
\int \int_{-\pi}^{+\pi} {dk_x^\prime dk_y^\prime \over 4\pi^2}
\displaystyle{\tanh(\epsilon_{k'}/2T)
\over 2\epsilon_{k'}
}\times
$$
\begin{equation}
\left[
\displaystyle{ {V_s\over \kappa_s^2+\eta_{k-k'-Q_s}}-
{V_c\over \kappa_c^2+\eta_{k-k'-Q_c}} }\right]
\Delta_{k'}
\label{bcs}
\end{equation}
where $\epsilon_k=\sqrt{\xi_k^2+\Delta_k^2}$, $\Delta_k$
is the gap parameter, and 
both the charge- and spin-induced static effective interactions 
in the Cooper channel have been considered, corresponding 
to an interaction in the particle-hole channel (\ref{eff}). 
A constructive
interference between the small-$q$ attraction associated to charge
fluctuations and the large-$q$ repulsion associated to spin fluctuations
yields both a high critical temperature and a gap parameter $\Delta_k$
with $d$-wave symmetry (Fig. 3).

The variation of the critical temperature with doping follows the
variation of the relevant energy scale in each region of the phase diagram.
In the overdoped (quantum disordered) region
$\kappa_c^2\sim x-x_c$ (at low temperatures)
and $T_c$  decreases rapidly
with increasing doping. In the quantum critical region around optimal doping
$\kappa_c^2\sim T$ and $T_c$ is almost doping independent.

In the underdoped region
new scales of energy appear. At $T_0(x)$,
which correspond to the mean-field ICDW critical temperature,
precursors of the stripe phase show up in the
reduction of spectral weight near the Fermi energy, i.e. a reduction
of the static spin susceptibility \cite{CHI}. At the same
time the damping of the AFM fluctuations is reduced and (almost) propagating
spin waves appear, with a natural crossover from a dynamical index $z=2$
to $z=1$. 
At lower temperatures, since
Boebinger's experiment \cite{boebinger}
suggests that superconductivity is hindering the formation
of a static CDW, we have to assume that the onset of (local) superconductivity
introduces a cut-off for critical fluctuations.
Thus, near the onset of the stripe phase $T_{CDW}(x)\lesssim T^*(x)$,
we take a mass term
$\kappa_c^2\simeq \max[\Delta_{max}(T),T-T_{CDW}(x)]$, where 
$\Delta_{max}$ is the maximum over $k$ of the (local) superconducting gap
$|\Delta_k|$. When introduced
into (\ref{bcs}), this dependence allows the (local) gap to survive up to
a temperature as high as $T^*$, even if the phase coherence, which
is necessary for bulk superconductivity, develops at a lower temperature
$T_c$. This produces a long pseudogap tail in the underdoped region, which, 
despite the crudeness of our approximations, displays the behavior of the
analogous quantity as measured in ARPES experiments
in underdoped Bi2212 \cite{harris}.

\section{Conclusions}
In this paper we 
briefly recapitulated the stripe-QCP scenario and presented some of its
consequences in the normal and SC states. Within
this scenario, the occurrence of a charge-ordering instability, only
hindered by the setting in of a SC phase (in this sense 
it would be more appropriate to speak about a ``missed-QCP'' 
scenario) provides the underlying mechanism ruling the physics of 
the SC cuprates. In particular, it gives rise to the 
formation of the observed stripe textures in these materials, to the 
non-Fermi liquid properties of the normal phase, to the main features
found in ARPES experiments, and to the strong 
pairing interaction. 
The most natural location for this QCP is optimum
doping where the strongest violation to the FL behavior and the 
highest critical temperature occur. Indeed the physical 
properties governed by the proximity to a QCP account for the 
ubiquitous {\it universal} behavior, observed near optimal doping
in all SC copper oxides.
This rationale is
missing in the theory of the AFM QCP or in the theory of the QCP associated to 
the coexistence of antiferromagnetism and superconductivity, which would 
also be located near the AF phase at low doping
or near the spin glass transition.

We conclude by remarking that 
the scenario of the stripe QCP near optimum doping,
hidden by the occurrence of the SC phase, shares a common origin 
(the Coulomb-frustrated phase separation) with the scenario 
proposed by Emery, Kivelson and coworkers, but relies on a distinct
mechanism. In this latter proposal, the anomalous normal properties 
stem from the marked one-dimensional character of the metallic 
stripe phase, and the pairing arises from the in-and-out pair 
hopping from the 1D stripes into the spin-gapped AF background. A 
related description of the stripe phase in terms of purely one-dimensional
strings has been also put forward by Zaanen \cite{zzz}. 
In our picture the non-FL character of the metal arises from the singular 
scattering by critical fluctuations near the QCP for the onset of the stripe 
phase. The fluctuations of the stripe texture also provide a strong 
pairing potential accounting for high critical temperatures.
We believe that this more two-dimensional physical 
description in terms fluctuations of the stripe texture
is closer to the reality, at least
for the optimal and overdoped systems, where the substantial 
metallic character of the systems is difficult to reconcile with the 
formation of strongly one-dimensional long-living stripe structures.  

{\bf Acknowledgements:} Part of this work was carried out with the financial 
support of the INFM, PRA 1996.   


\begin{figure}[htbp]   
    \begin{center}
       \setlength{\unitlength}{1truecm}
       \begin{picture}(5.0,15.0)    
          \put(-6.0,-8.0){\epsfbox{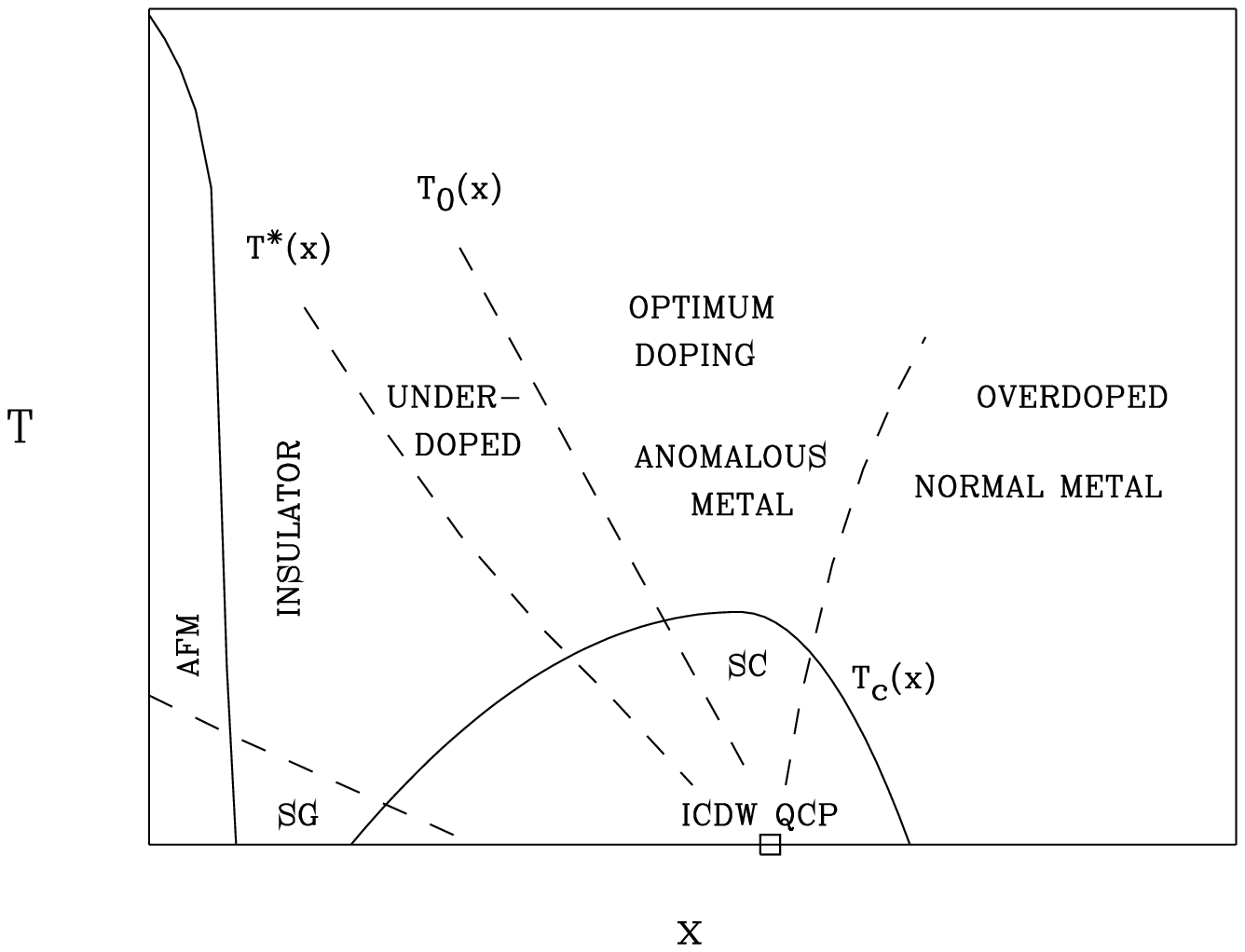}}   
       \end{picture}
    \end{center}
    \caption{Generic phase diagram for a SC copper oxide. The meaning of the
    temperature scales is explained in the text.}
    \protect\label{fig1}
\end{figure}
\begin{figure}[htbp]   
    \begin{center}
       \setlength{\unitlength}{1truecm}
       \begin{picture}(5.0,15.0)    
          \put(-6.0,-8.0){\epsfbox{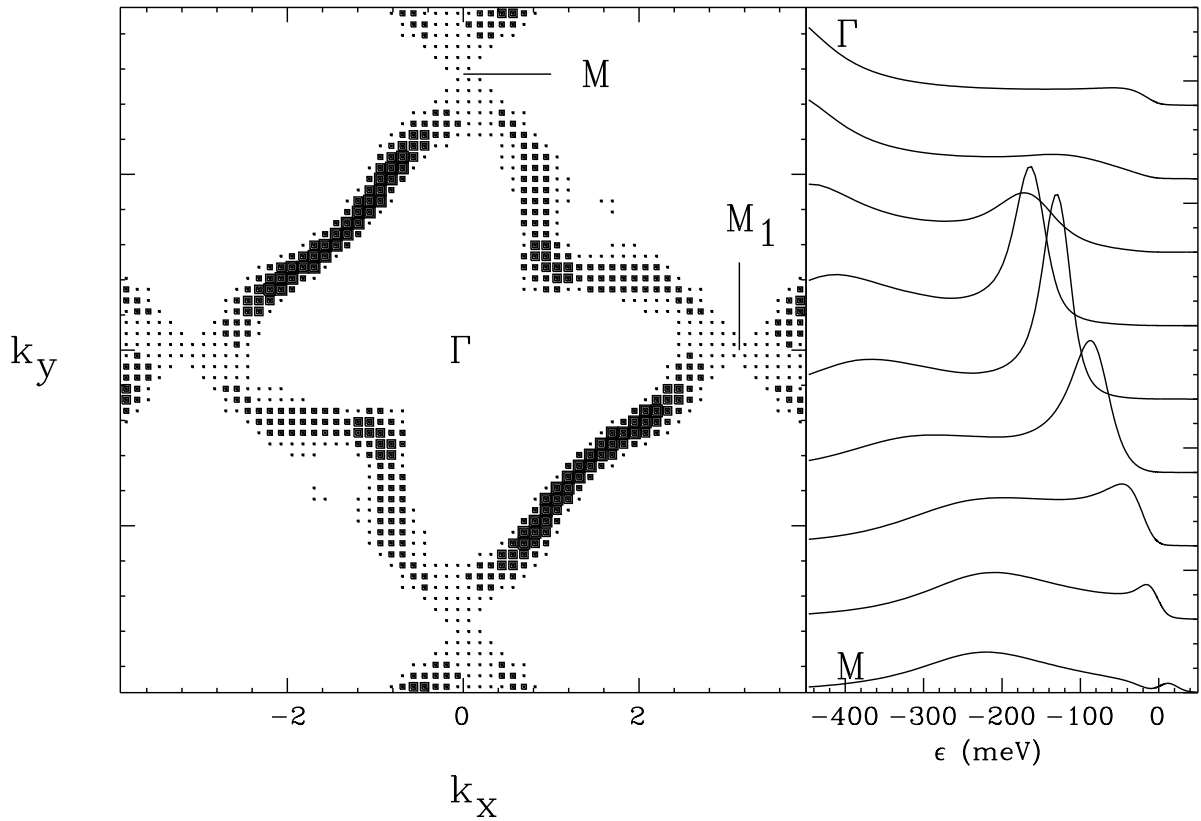}}   
       \end{picture}
    \end{center}
    \caption{Left: $k$-space distribution of the low-laying spectral weight
    in the case of an effective electron-electron interaction mediated
    by both charge and spin fluctuations. The values of the parameters
    are given in the text.}
    \protect\label{fig2}
\end{figure}
\begin{figure}[htbp]
\protect
\centerline{\psfig{figure=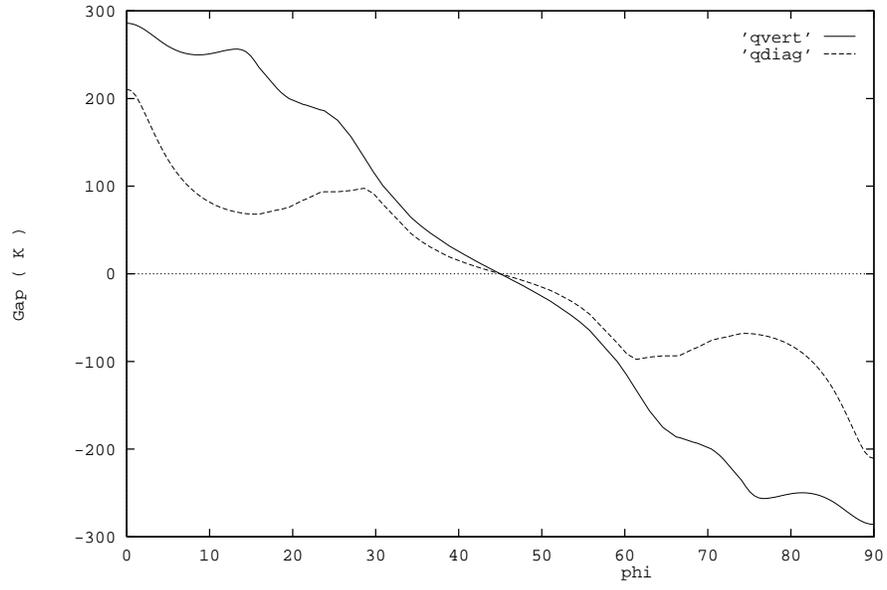,width=12cm,angle=-90}}
    \caption{Gap parameter on the FS as a function of the polar angle
    from the point $(\pi,\pi)$.}
    \label{fig3}
\end{figure}
\end{document}